\documentclass[a4paper,fleqn]{cas-sc}

\usepackage[numbers]{natbib}
\usepackage[utf8]{inputenc}

\def\tsc#1{\csdef{#1}{\textsc{\lowercase{#1}}\xspace}}
\tsc{WGM}
\tsc{QE}

\begin{document}
\let\WriteBookmarks\relax
\def\floatpagepagefraction{1}
\def\textpagefraction{.001}

\shorttitle{Electrolyte-Dependent Structure–Transport Relationships in Electrodeposited Prussian Blue Analogue Thin Films}    

\shortauthors{Garcia et al.}  

\title [mode=title]{Electrolyte-Dependent Structure–Transport Relationships in Electrodeposited Prussian Blue Analogue Thin Films}  

%

\author[1]{Larissa de O. Garcia}
\cormark[1]
\ead{larissa.de.oliveira.garcia@whz.de}
\ead[url]{ORCID: https://orcid.org/0000-0003-0267-6488}

\author[1]{Michael Pohlitz}

\author[2]{Mohammed F. Kalady}

\author[1]{Christian K. Müller}


\affiliation[1]{
organization={Faculty of Physical Engineering/Computer Sciences, University of Applied Sciences Zwickau},
city={Zwickau},
postcode={08056},
country={Germany}
}

\affiliation[2]{
organization={Leibniz Institute for Solid State and Materials Research Dresden (IFW Dresden)},
city={Dresden},
postcode={01069},
country={Germany}
}

----------------------------------------
\cortext[1]{Corresponding author. Tel.: +49 375 536 1510}


\begin{abstract}
Understanding how electrolyte composition influences the structural factors governing charge storage in Prussian Blue analogues (PBAs) requires clarifying the coupled effects of ion desolvation, structural disorder, and transport pathway connectivity. Here, we demonstrate that electrolyte identity induces structural disorder, which correlates with transport dispersion measured under identical electrochemical conditions.

To establish this relationship, Fe-, Co-, and Ni-based hexacyanoferrate thin films were electrodeposited from chloride electrolytes (KCl, NaCl, NH$_4$Cl, and LiCl), enabling a comparative investigation of structure-transport correlations across multiple length scales.

Although all systems crystallize in the cubic PBA structure, pronounced electrolyte-dependent variations in lattice parameters, defect concentration, and local coordination environments are observed. Na$^+$ induces lattice expansion but simultaneously promotes vacancy formation, microstrain, and structural heterogeneity, whereas K$^+$ and NH$_4^+$ produce more structurally coherent frameworks with reduced disorder. Raman spectroscopy reveals that increasing structural disorder broadens the distribution of local coordination environments, which is accompanied by increasingly dispersed electrochemical behavior.

Electrochemical impedance spectroscopy reveals that charge storage is governed by coupled ion-electron transport processes, with the impedance increasing by more than one order of magnitude from FeHCF to NiHCF. Importantly, the constant phase element (CPE) exponent decreases systematically with increasing Raman band broadening, revealing a strong correlation between structural disorder and transport dispersion.

These results demonstrate that electrolyte identity controls not only lattice dimensions but, more importantly, the organization and connectivity of defect networks. Consequently, optimal electrochemical performance arises from a balance between structural coherence and accessible transport pathways rather than from lattice expansion alone.

\end{abstract}

\begin{graphicalabstract}
\centering
\includegraphics[width=\textwidth]{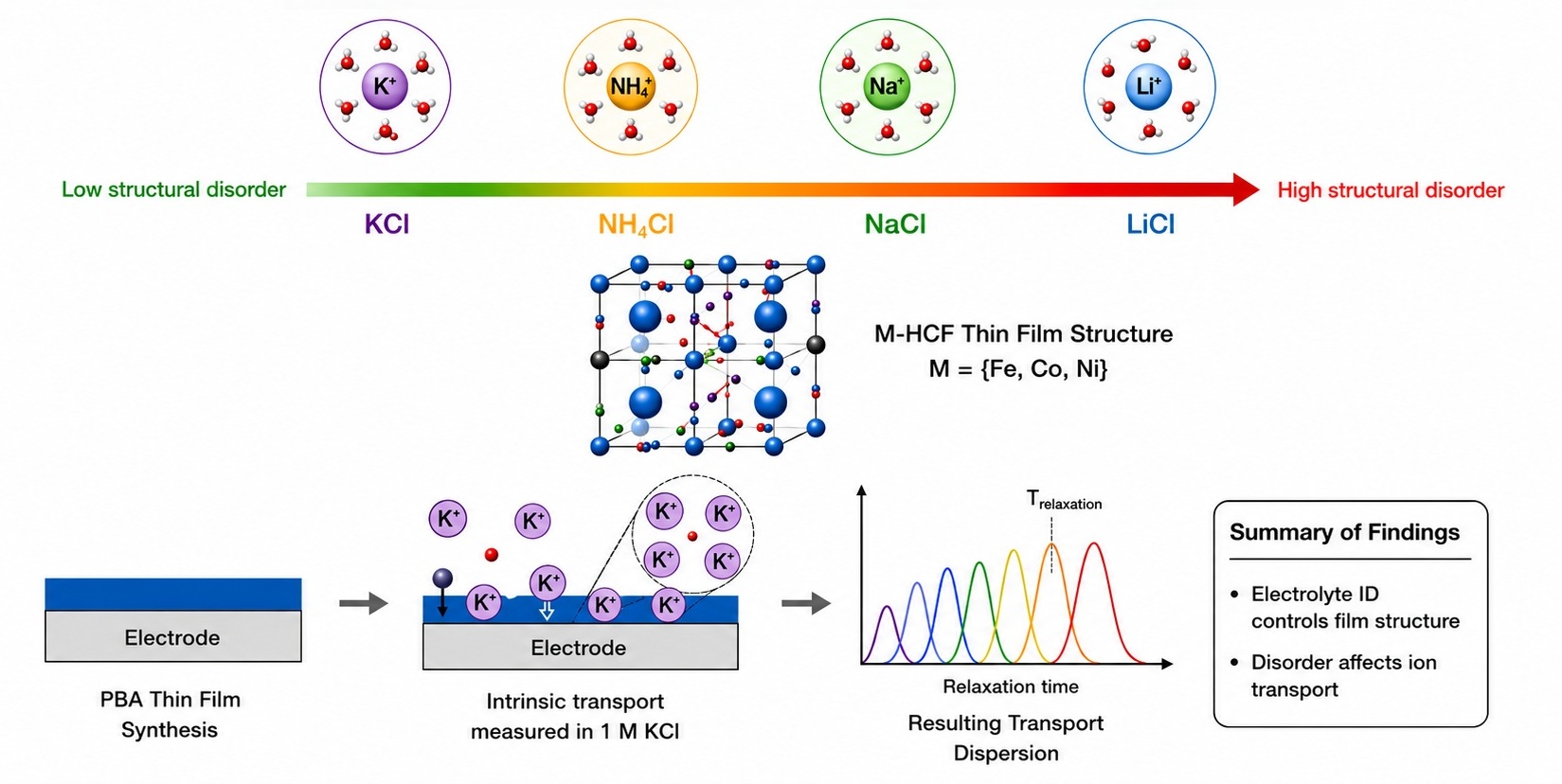}
\end{graphicalabstract}

\begin{highlights}

\item Electrolyte identity strongly influences structural disorder in PBAs
\item Structural disorder broadens relaxation times and limits ion transport
\item Raman, EIS, and CV reveal correlated structure--transport behavior
\item Charge storage arises from distributed ion--electron transport processes
\item Structural disorder strongly influences transport connectivity

\end{highlights}

\maketitle

\section{Introduction}

Prussian Blue (PB) and its analogues (PBAs) are open-framework coordination compounds with the general formula $A_xM[M'(CN)_6]_y \cdot nH_2O$, crystallizing in a face-centered cubic lattice that contains large interstitial cavities capable of accommodating electrolyte cations \cite{Wang2017,Mondal2026}. Their electrochemical activity arises from coupled ion-electron transfer processes, in which faradaic redox reactions at transition-metal centers are charge-compensated by reversible cation insertion within the framework \cite{Itaya1982,Feldman1985}. Owing to this mechanism, PBAs have emerged both as model systems for investigating ion insertion in defect-rich, hydrated materials and as promising cathode materials for aqueous alkali-ion batteries \cite{Pasta2012,Wessells2011,Wu2021,Xiao2024,Wang2020}. Electrodeposited PB and related PBA thin films have also been investigated for electrochemical devices and resistive-switching or memristive applications, where film formation, defect structure, and electrochemical operation are closely coupled \cite{Avila2020,Faita2022,Avila2022,Avila2024PB,Avila2024Perylene,Avila2025,Cantudo2026,Avila2026Nanoscale}.

Beyond their average crystal structure, PBAs exhibit a high degree of structural complexity, including vacancy defects, coordinated water molecules, and lattice distortions that interact across multiple length scales \cite{Pasta2012,Cattermull2021,Simonov2020,Shen2021}. Together, these structural features define a correlated defect landscape that governs both the physical and electrochemical properties of the framework \cite{Rodriguezetal2021,Grandjean2016,Itaya1986}. In particular, cooperative phenomena such as Jahn-Teller distortions and defect-induced strain play important roles in determining lattice stability, ion accessibility, and transport characteristics \cite{Pasta2016,Moritomo2018}.

Charge storage in PBAs differs fundamentally from that in conventional intercalation materials \cite{Sadiq2026,Iyyappan2026}. It involves a multi-step process comprising electron transfer at the electrode interface, cation desolvation at the electrolyte boundary, interfacial ion transport, and solid-state diffusion within the framework \cite{Vorotyntsev2008,Augustyn2013,Come2014}. These processes occur simultaneously and are intrinsically interconnected, giving rise to a distributed kinetic response rather than a single well-defined charge-transfer step. Furthermore, nanoscale studies have demonstrated that electronic conductivity in PBAs is spatially heterogeneous and strongly dependent on the local structural environment, further highlighting the intimate relationship between ionic and electronic transport \cite{Ishizaki2013,Hurlbutt2018}.

From a thermodynamic perspective, the free energy of ion insertion may be expressed as contributions from the intrinsic redox potential of the host framework, the desolvation energy of the inserting cation, electrostatic interactions within the lattice, and the mechanical strain associated with structural distortion during insertion \cite{Marcus1994,Cheng2017}. In aqueous electrolytes, desolvation plays a particularly important role: Li$^+$ exhibits large insertion barriers because of its high hydration enthalpy, whereas larger cations such as K$^+$ and NH$_4^+$ are more readily desolvated but may induce more pronounced lattice strain during insertion \cite{Wu2018}.

Despite extensive investigation, the relationship between lattice expansion, defect formation, and electrochemical performance remains unresolved. Some studies attribute improved transport kinetics to increased lattice parameters and enhanced ion accessibility \cite{Pasta2012,Gao2024}. In contrast, other studies report that structural disorder and vacancy formation hinder ion transport despite lattice expansion \cite{Cattermull2021,Wu2024}. One possible explanation for these apparently contradictory observations is the influence of coordinated water molecules and local structural distortions. For example, You \textit{et al.} demonstrated that zeolitic water occupying vacancy sites acts as a steric barrier that reduces the theoretical accessibility of the framework \cite{You2015}. Likewise, Traducci \textit{et al.} showed that, for strongly hydrated ions such as Li$^+$, interactions with local defect environments and coordinated water molecules dominate the insertion barrier more significantly than the average lattice size \cite{Traducci2020}. These observations indicate that lattice expansion alone is insufficient to describe ion transport in PBAs and suggest that the organization of defect networks and their influence on transport continuity must also be considered \cite{Zhou2024}.

This issue is particularly relevant for electrodeposited PBA thin films, where the electrolyte used during synthesis can directly influence vacancy concentration, coordinated water content, and interstitial ion incorporation \cite{Garcia2025,Pohlitz2022,Avila2020,Avila2025,Avila2026Nanoscale}. Recent studies of electrodeposited PB-based devices further show that device-to-device variability, conductance behavior, and nanoscale resistive switching can be strongly affected by the film microstructure and the electrochemical environment \cite{Avila2025,Cantudo2026,Avila2026Nanoscale}. Although electrolyte effects are often interpreted primarily in terms of ionic radius and desolvation energy, their influence on defect organization and the resulting transport behavior has not been systematically investigated.

In this work, we show that electrolyte identity strongly influences not only lattice parameters and thermodynamic insertion properties but, more importantly, the structural disorder that governs transport in electrodeposited PBAs. Iron hexacyanoferrate (FeHCF) is employed as a model system because of its well-defined redox chemistry, allowing structural and kinetic contributions to be examined independently, while cobalt (CoHCF) and nickel (NiHCF) analogues provide a systematic framework for investigating the influence of transition-metal-dependent structural disorder.

By combining X-ray diffraction (XRD), Raman spectroscopy, scanning electron microscopy (SEM), energy-dispersive X-ray spectroscopy (EDS), cyclic voltammetry (CV), scan-rate analysis, and electrochemical impedance spectroscopy (EIS), we establish direct correlations between structural disorder, transport dispersion, and electrochemical response.

The results show that electrolyte-induced structural disorder governs the distribution of local coordination environments and electrochemical relaxation times, ultimately determining the transport behavior of the framework. These findings identify electrolyte selection as an effective strategy for controlling defect organization and transport limitations in PBAs, providing a mechanistic framework for understanding electro-chemo-mechanical behavior in cyanide-bridged open-framework materials.

\section{Experimental}

\subsection{Electrodeposition of PBA Thin Films}

Prussian Blue analogue (PBA) thin films were synthesized by potentiostatic electrodeposition using a conventional three-electrode electrochemical cell connected to an Ivium CompactStat potentiostat (Ivium Technologies, The Netherlands). A platinum plate served as the counter electrode, while a saturated calomel electrode (SCE) was used as the reference electrode. Au (50 nm)/Cr (5 nm)-coated silicon substrates (11.5 mm $\times$ 15 mm) were used as working electrodes. The substrates were prepared by electron-beam evaporation under high-vacuum conditions, in which a 5 nm Cr adhesion layer was deposited prior to deposition of the 50 nm Au film.

Electrodeposition was carried out potentiostatically at 0.30 V versus SCE until a total charge of 50 mC was reached. The effective deposition area ($\sim$0.5 cm$^2$) was defined using an insulating adhesive mask to ensure reproducible film geometry.

The deposition electrolyte consisted of 1.0 M supporting electrolyte (KCl, NaCl, LiCl, or NH$_4$Cl), 0.25 mM K$_3$[Fe(CN)$_6$], and 0.25 mM metal chloride precursor (FeCl$_3\cdot$6H$_2$O, CoCl$_2\cdot$6H$_2$O, or NiCl$_2\cdot$6H$_2$O). The solution pH was adjusted to approximately 2 using hydrochloric acid to ensure the stability of the ferricyanide precursor during electrodeposition.

The electrodeposition protocol followed procedures previously established for electrodeposited Prussian Blue and related electrochemical thin-film systems \cite{Avila2020,Garcia2025,Garcia2026,Quispe2021}, with the supporting-electrolyte cation being the only experimental variable. After deposition, the films were rinsed thoroughly with deionized water to remove residual electrolyte species and subsequently dried under ambient conditions. Throughout this work, the samples are designated according to the transition-metal precursor used during electrodeposition as FeHCF, CoHCF, and NiHCF.

\subsection{Structural and Morphological Characterization}

The crystal structure of the electrodeposited PBA thin films was investigated by X-ray diffraction (XRD) using a PANalytical X'Pert MRD diffractometer equipped with Co K$\alpha$ radiation ($\lambda = 1.78896$~\AA). Diffraction patterns were collected in Bragg--Brentano geometry over a 2$\theta$ range of 10--40$^\circ$.

Surface morphology and cross-sectional microstructure were examined by field-emission scanning electron microscopy (FEG-SEM, TESCAN CLARA, Brno, Czech Republic). Elemental composition was determined by energy-dispersive X-ray spectroscopy (EDS, Oxford Instruments) operated at an accelerating voltage of 20 kV.

Raman spectra were acquired using a confocal Raman microscope (WITec RISE, Ulm, Germany) equipped with a 532 nm excitation laser and a Zeiss LD EC Epiplan-Neofluar 100$\times$/0.75 objective. The laser power was limited to 0.4 mW to minimize local heating and avoid laser-induced modification of the films. Spectra were collected using a 600 grooves mm$^{-1}$ grating, an integration time of 0.7 s, and 100 accumulations. All Raman measurements were performed under identical acquisition conditions.

\subsection{Electrochemical Measurements}

Electrochemical characterization was performed using the same three-electrode cell described above and the Ivium CompactStat potentiostat. The use of electrodeposition-derived PBA films in electrochemical and nanoscale resistive-switching studies has been reported previously, including studies linking K-ion intercalation to local electrical switching behavior \cite{Avila2020,Faita2022,Avila2022,Avila2026Nanoscale}.

Cyclic voltammetry (CV) measurements were carried out over the potential range of 0.0--0.6 V versus SCE, following the broader use of electrochemical routes for thin-film preparation and characterization \cite{Quispe2021}. This potential window was selected to avoid oxidation of the Au working electrode in chloride-containing electrolytes at higher anodic potentials, thereby minimizing substrate-related faradaic contributions and enabling direct comparison among the investigated PBA thin films. Unless otherwise stated, the supporting electrolyte used during electrochemical characterization was identical to that employed during electrodeposition.

To evaluate the intrinsic transport behavior of the different PBA frameworks, scan-rate measurements were performed exclusively on films electrodeposited in 1 M KCl and subsequently characterized in the same electrolyte. Measurements were carried out at scan rates of 10, 25, 50, and 100 mV s$^{-1}$.

The anodic and cathodic peak currents were obtained directly from the corresponding peak maxima without baseline correction or peak deconvolution and were subsequently used to determine the apparent $b$-values.

Electrochemical impedance spectroscopy (EIS) measurements were also performed on films deposited and characterized in 1 M KCl. Spectra were recorded at the formal potential of the corresponding redox couple using a sinusoidal perturbation of 5 mV over the frequency range from 100 kHz to 0.1 Hz.

Different film thicknesses were intentionally employed according to the characterization technique. Thin films (approximately 200 nm) were used for electrochemical measurements to minimize diffusion limitations and facilitate evaluation of the intrinsic transport properties. Thicker films (approximately 0.6--1.7 $\mu$m) were used for XRD, Raman spectroscopy, SEM, and EDS analyses to improve signal quality while minimizing substrate contributions. Since both film types were prepared using the same electrodeposition protocol, the observed structural and electrochemical differences are attributed primarily to the influence of the supporting electrolyte on the resulting PBA framework rather than to differences in the deposition procedure.

\section{Results and Discussion}

\subsection{Electrolyte-Dependent Structural Evolution}

\begin{figure}[H]
\centering
\includegraphics[width=\textwidth]{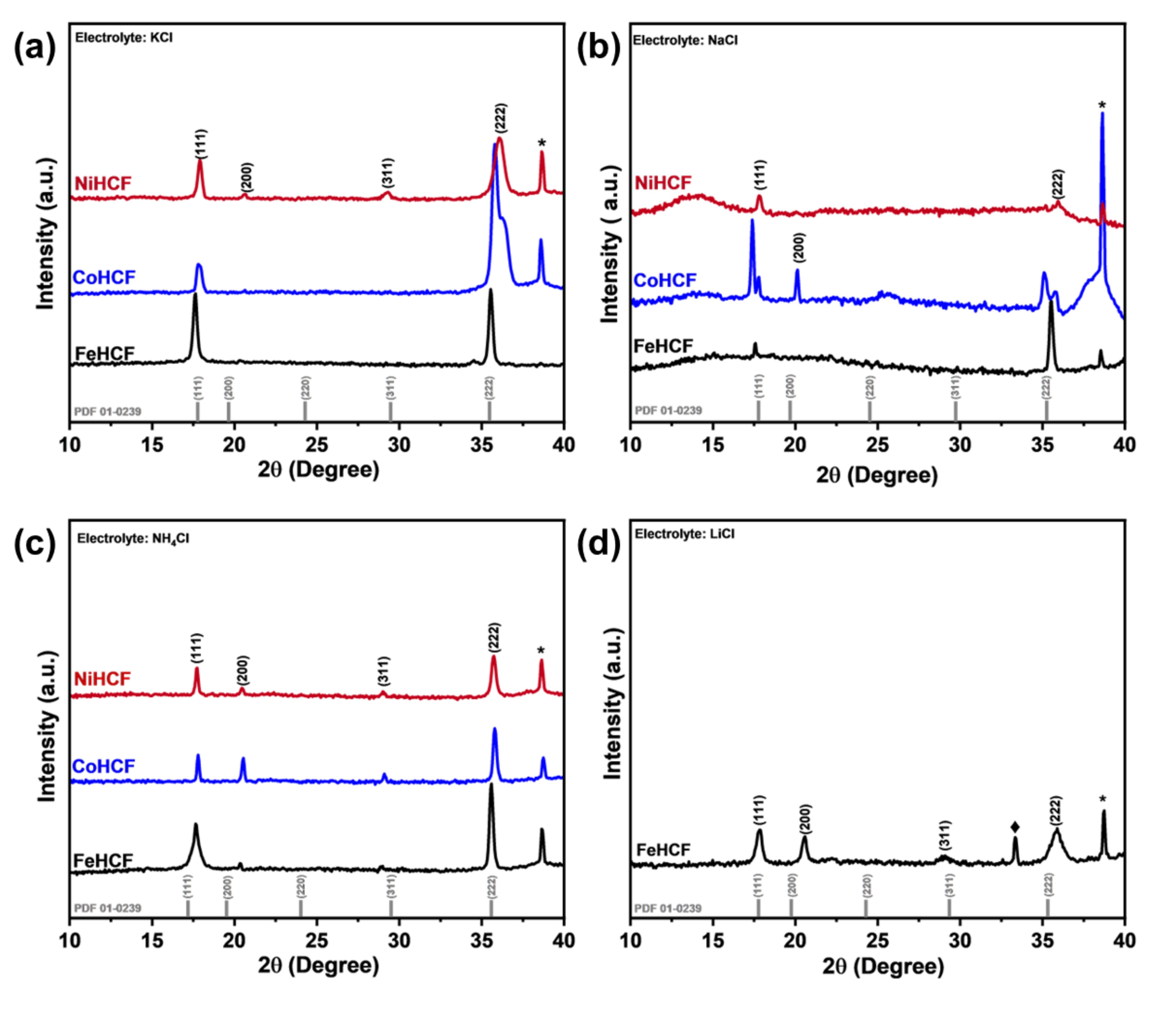}
\caption{
X-ray diffraction patterns of FeHCF, CoHCF, and NiHCF films electrodeposited in different supporting electrolytes using Co K$\alpha$ radiation. Additional features marked by (*) are attributed to surface oxide formation, whereas the weak reflection indicated by ($\diamond$) at approximately $2\theta \approx 33^\circ$ remains unassigned.
}
\label{fig:xrd}
\end{figure}

To evaluate the influence of the supporting electrolyte on the crystal structure of the electrodeposited Prussian Blue analogue (PBA) thin films, X-ray diffraction (XRD) patterns were collected using Co K$\alpha$ radiation ($\lambda = 1.789$~\AA) (Figure~\ref{fig:xrd}). Reference peak positions were obtained from PDF 01-0239 and converted from Cu K$\alpha$ to Co K$\alpha$ radiation for phase identification.

All samples crystallize in the cubic Prussian Blue analogue structure (space group Fm$\bar{3}$m), exhibiting the characteristic (111), (200), (220), (311), and (222) reflections. No secondary crystalline phases associated with the PBA framework are detected, confirming phase purity under all investigated deposition conditions. Nevertheless, pronounced electrolyte-dependent variations in peak position, intensity, and width reveal significant differences in lattice parameter and structural coherence.

Films deposited in KCl exhibit sharp, well-defined diffraction peaks, indicating a comparatively ordered framework with low structural disorder. The calculated lattice parameters agree well with our previous reports obtained under similar electrodeposition conditions, supporting the use of KCl as a reference electrolyte for structurally coherent PBA thin films \cite{Garcia2025,Garcia2026}.

\begin{table}[H]
\centering
\caption{Lattice parameters ($a$) calculated from the (111) reflection assuming a cubic Fm$\bar{3}$m structure.}
\begin{tabular}{lccc}
\hline
Electrolyte & FeHCF (Å) & CoHCF (Å) & NiHCF (Å) \\
\hline
KCl      & 10.12 & 10.01 & 9.99 \\
NaCl     & 10.16 & 10.22 & 9.98 \\
NH$_4$Cl & 10.08 & 10.03 & 10.08 \\
LiCl     & 10.03 & --    & --    \\
\hline
\end{tabular}
\label{tab:lattice_parameters}
\end{table}

In contrast, NaCl-derived films exhibit a systematic shift of the diffraction peaks toward lower $2\theta$ values, indicating lattice expansion (Table~\ref{tab:lattice_parameters}). This effect is most pronounced for CoHCF, whose lattice parameter increases from 10.01 to 10.22~\AA. Simultaneously, CoHCF and NiHCF display marked peak broadening, indicative of increasing microstrain and reduced long-range crystallographic order. The diffraction pattern of NaCl-derived NiHCF becomes particularly weak and broadened, approaching an amorphous-like character. Similar structural degradation has previously been associated with high concentrations of lattice defects and structural distortions in Prussian Blue analogues \cite{Sathishkumar2025,Feng2026}.

The combined peak shift and broadening indicate that lattice expansion is accompanied by increasing structural disorder rather than improved crystallinity. In PBAs, such behavior has frequently been associated with the incorporation of [Fe(CN)$_6$] vacancies, coordinated water molecules, and the resulting microstrain, all of which disrupt framework connectivity and reduce long-range order \cite{Cafun2013,Takachi2013,Nie2025}.
LiCl-derived FeHCF exhibits a slight shift toward higher $2\theta$ values, corresponding to a contracted lattice. This behavior is consistent with the limited incorporation of Li$^+$, whose high hydration energy imposes a substantial desolvation barrier prior to insertion \cite{Wang2026Li}. Under the investigated conditions, CoHCF and NiHCF could not be reproducibly electrodeposited in LiCl, further indicating that Li$^+$ does not effectively support framework formation.

Films deposited in NH$_4$Cl exhibit intermediate structural characteristics, combining relatively small lattice distortions with moderate crystallinity. This behavior is consistent with the distinct interaction of NH$_4^+$ with the PBA framework, where hydrogen-bond-mediated stabilization has been proposed to partially suppress structural distortion \cite{Maiti2026,Geng2026}.

The comparatively weak higher-order reflections are attributed to the combined effects of thin-film geometry, preferred orientation during electrodeposition, structural disorder, and fluorescence background arising from the use of Co K$\alpha$ radiation \cite{He2003}.

Overall, the XRD results demonstrate that the supporting electrolyte strongly influences the long-range structure of electrodeposited PBAs by modifying lattice dimensions, crystallographic coherence, and defect accommodation. While KCl promotes the formation of comparatively ordered frameworks, NaCl favors lattice expansion accompanied by increasing structural disorder, indicating that larger lattice parameters do not necessarily correspond to improved structural quality.

\subsection{Local Coordination and Defect Chemistry}

\begin{figure}[h]
\centering
\includegraphics[width=0.9\textwidth]{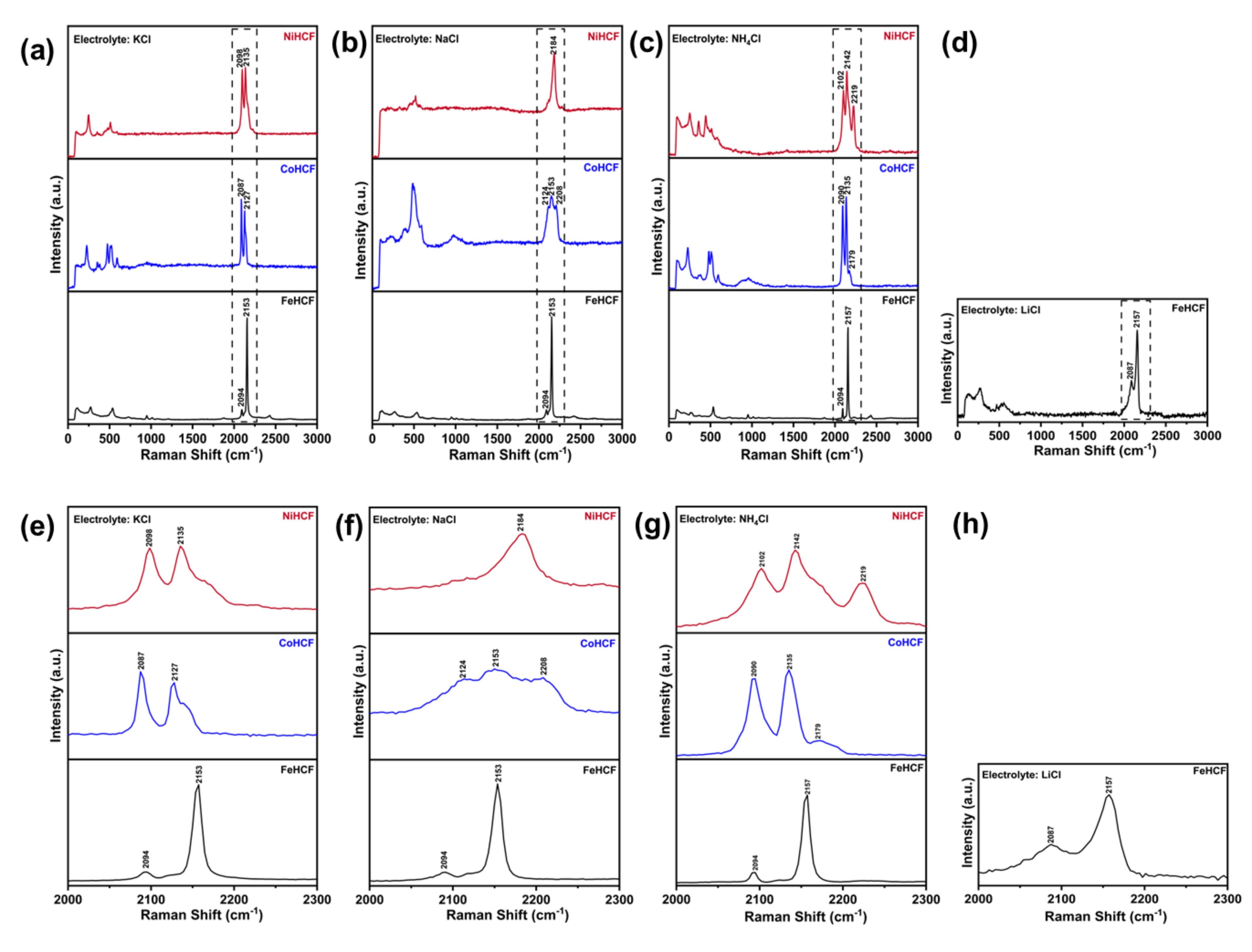}
\caption{
Raman spectra of FeHCF, CoHCF, and NiHCF thin films electrodeposited in different supporting electrolytes. Electrolyte-dependent variations in peak position and full width at half maximum (FWHM) are observed within the CN stretching region (2000--2200 cm$^{-1}$).
}
\label{fig:raman}
\end{figure}

While X-ray diffraction probes the average long-range crystallographic structure, Raman spectroscopy is considerably more sensitive to local coordination environments and short-range disorder. It was therefore employed to investigate the local bonding configuration of the Fe--C$\equiv$N--M framework (Figure~\ref{fig:raman}), providing complementary information on the local electronic structure of the electrodeposited Prussian Blue analogue (PBA) thin films \cite{Giorgetti2012}.

The CN stretching region (2000--2200 cm$^{-1}$) is highly sensitive to the local coordination environment of cyanide-bridged frameworks. The $\nu$(CN) frequency depends on the electronic structure of the Fe--C$\equiv$N--M linkage and is affected by the transition-metal identity, oxidation state, ligand field, local symmetry, hydration, and metal-to-ligand $\pi$-backbonding. Consequently, frequency shifts reflect changes in the local electronic environment, whereas linewidth broadening indicates an increasing distribution of nonequivalent coordination environments \cite{Nakamoto2009,Kettle2011}.

KCl-derived films exhibit comparatively narrow Raman bands with well-resolved CN stretching modes (Table~\ref{tab:Raman_CN}), indicating relatively homogeneous local coordination environments and confirming their superior structural coherence.

In contrast, NaCl-derived films show pronounced band broadening, particularly for CoHCF, where the FWHM reaches approximately 269 cm$^{-1}$. Simultaneous shifts in the CN stretching frequencies indicate substantial modifications of the local electronic environment. These features are characteristic of increased structural disorder and have previously been associated with higher concentrations of [Fe(CN)$_6$] vacancies, coordinated water molecules, local symmetry breaking, and lattice distortions in Prussian Blue analogues, although these contributions cannot be distinguished individually in the present study \cite{Simonov2020,Bostrom2022}.

For NaCl-derived NiHCF, the CN stretching region collapses into a single broad band centered near 2183 cm$^{-1}$, indicating a wide distribution of local bonding environments and a pronounced loss of coordination homogeneity. This behavior is consistent with the diffuse XRD pattern observed for this sample and suggests extensive short-range disorder.

\begin{table}[h]
\centering
\caption{CN stretching frequencies ($\nu$) and full width at half maximum (FWHM) obtained from Raman spectra of PBA thin films prepared in different supporting electrolytes. Increasing FWHM values indicate greater local structural disorder and a broader distribution of coordination environments.}
\label{tab:Raman_CN}
\begin{tabular}{llcccc}
\hline
\textbf{Electrolyte} & \textbf{Sample} & $\nu_1$ & $\nu_2$ & FWHM$_1$ & FWHM$_2$ \\
& & (cm$^{-1}$) & (cm$^{-1}$) & (cm$^{-1}$) & (cm$^{-1}$)\\
\hline
\multirow{3}{*}{KCl}
& FeHCF & 2094.7 & 2157.3 & 20.0 & 13.4 \\
& CoHCF & 2087.3 & 2127.9 & 18.8 & 33.6 \\
& NiHCF & 2098.4 & 2135.3 & 33.3 & 55.5 \\
\hline
\multirow{3}{*}{NaCl}
& FeHCF & 2091.0 & 2153.7 & 30.9 & 14.3 \\
& CoHCF & 2150.0 & 2208.6 & 66.1 & 269.2 \\
& NiHCF & -- & 2183.0 & -- & 137.7 \\
\hline
\multirow{3}{*}{NH$_4$Cl}
& FeHCF & 2094.7 & 2157.3 & 10.1 & 11.1 \\
& CoHCF & 2094.7 & 2135.3 & 21.9 & 21.1 \\
& NiHCF & 2102.1 & 2142.6 & 34.4 & 52.8 \\
\hline
LiCl
& FeHCF & 2087.3 & 2157.3 & 62.4 & 34.0 \\
\hline
\end{tabular}
\end{table}

LiCl-derived FeHCF exhibits moderate band broadening together with relatively small frequency shifts, consistent with the limited Li$^+$ insertion expected from its strong hydration shell and high desolvation energy \cite{Bie2022,Wessells2012}. In contrast, NH$_4$Cl-derived films display comparatively narrow Raman bands, suggesting relatively uniform local coordination environments, which may result from hydrogen-bond-mediated stabilization involving NH$_4^+$ species within the framework \cite{Song2018Joule}.

An additional CN stretching component appears in NH$_4$Cl-derived NiHCF and NaCl-derived CoHCF. The coexistence of multiple CN stretching modes indicates that the Fe--C$\equiv$N--M linkage is distributed over nonequivalent local environments rather than a single coordination geometry. Similar spectral splitting has been attributed to local symmetry breaking, mixed Fe oxidation states, vacancy-induced distortions, coordinated water, and interactions between interstitial species and the cyanide framework \cite{Simonov2020,Bostrom2022,Song2018Joule}. Although the present data do not allow these effects to be separated individually, the spectral evolution clearly indicates increasing local structural heterogeneity.

The Raman results also demonstrate that long-range crystallographic order does not necessarily imply local electronic homogeneity. Samples exhibiting relatively sharp diffraction peaks may still display multiple CN stretching contributions arising from distinct short-range coordination environments, highlighting the complementary nature of Raman spectroscopy and XRD.

These structural trends closely parallel the electrochemical response discussed in the following sections. Samples exhibiting broader Raman bands generally present lower apparent $b$-values, lower CPE exponents, and increasingly non-ideal voltammetric behavior, indicating progressively more heterogeneous transport pathways. Since the CPE exponent is influenced by multiple factors, including surface roughness, porosity, structural disorder, and current distribution, it is interpreted here as an effective descriptor of transport heterogeneity rather than direct evidence of a unique relaxation-time distribution \cite{CordobaTorres2015,Hirschorn2010}.

Overall, Raman spectroscopy reveals that the supporting electrolyte strongly influences the local coordination chemistry of electrodeposited PBA thin films. Combined with the XRD results, it demonstrates a progressive transition from relatively homogeneous frameworks to increasingly disordered local coordination environments, providing the structural basis for the transport behavior discussed in the following electrochemical sections.

\subsection{Morphology, Growth Behavior, and Composition}

\begin{figure}[h]
\centering
\includegraphics[width=0.9\textwidth]{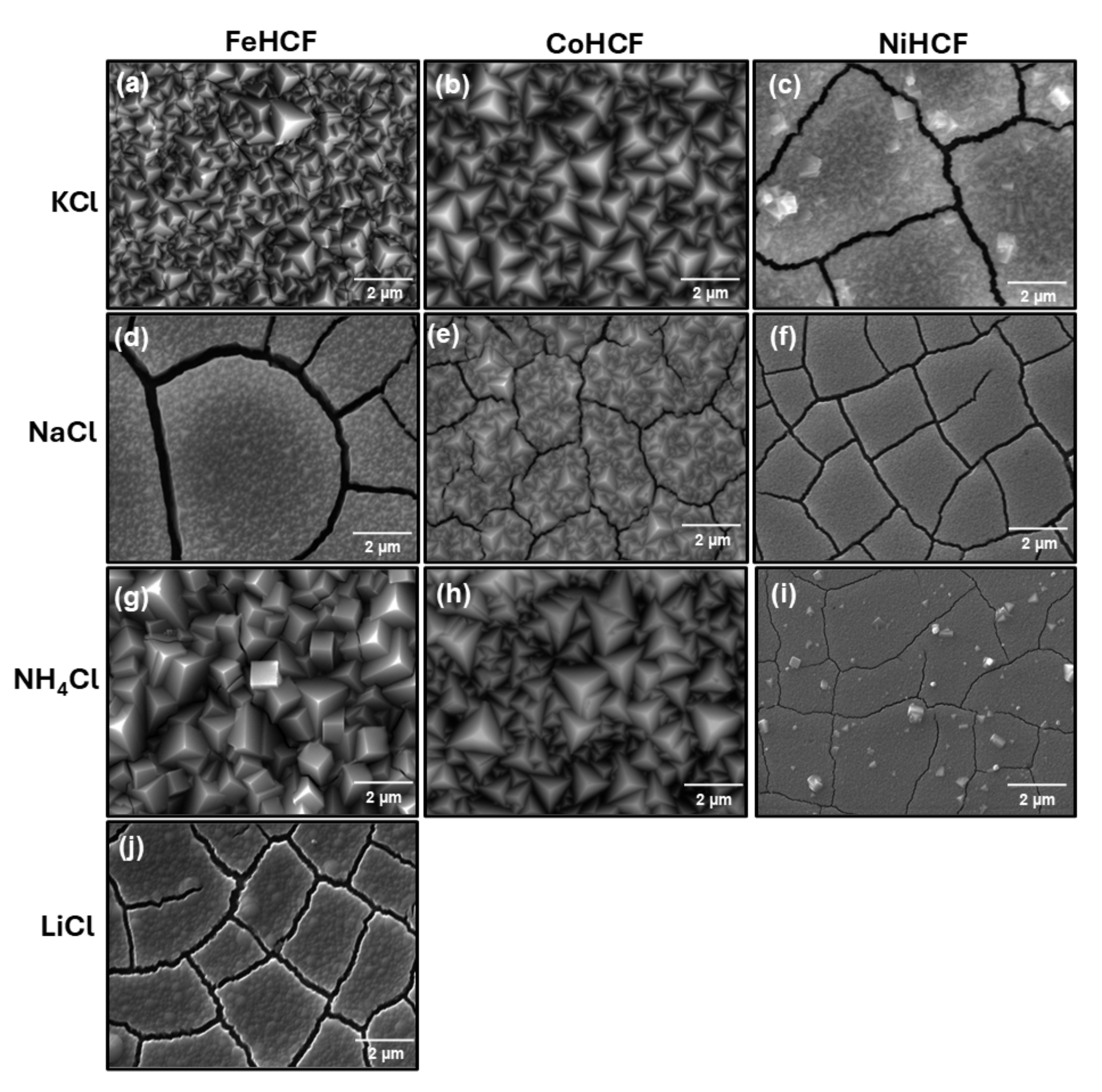}
\caption{
Top-view SEM images of PBA thin films electrodeposited using different supporting electrolytes.
}
\label{fig:sem}
\end{figure}

Scanning electron microscopy (SEM) was employed to investigate the influence of the supporting electrolyte on the morphology of the electrodeposited PBA thin films. As shown in Figure~\ref{fig:sem}, electrolyte-dependent structural variations are accompanied by pronounced changes in surface morphology, indicating that modifications in crystallographic order and local coordination directly affect mesoscale film growth.

KCl-derived films exhibit comparatively dense and continuous morphologies composed of well-defined crystalline grains with limited cracking, indicating relatively homogeneous growth and good mechanical integrity. Among the investigated materials, FeHCF and CoHCF prepared in KCl display the most stable morphologies.

In contrast, NaCl-derived films exhibit extensive cracking and fragmentation regardless of composition, consistent with internal stress accumulation associated with lattice expansion, defect incorporation, and heterogeneous strain distribution within the framework \cite{Mukhopadhyay2014}. The effect is particularly pronounced for CoHCF and NiHCF, which also exhibit the largest structural disorder according to XRD and Raman spectroscopy.

Interestingly, additional crack formation was observed for NaCl-derived FeHCF and CoHCF during SEM examination under high vacuum. This behavior suggests that these films retain comparatively large amounts of coordinated or trapped water, which are progressively removed during vacuum exposure, generating dehydration-induced stress and further mechanical fragmentation \cite{Maddar2023,Sada2024}. Together with the increased oxygen content determined by EDS, these observations indicate that NaCl promotes the formation of highly hydrated and defect-rich frameworks.

Among the investigated materials, CoHCF appears particularly sensitive to electrolyte-induced structural modification. The pronounced lattice expansion, broad Raman bands, and severe surface cracking observed for NaCl-derived CoHCF suggest enhanced incorporation of coordinated water, vacancies, and local lattice strain relative to FeHCF and NiHCF.

NiHCF exhibits a distinct mechanical response. Extensive cracking is observed irrespective of the supporting electrolyte, indicating an intrinsic susceptibility of the Ni-based framework to residual stresses generated during electrodeposition. Unlike NaCl-derived FeHCF and CoHCF, which develop additional cracks during SEM observation, NiHCF already exhibits dense crack networks immediately after deposition, suggesting that mechanical instability develops primarily during film growth rather than subsequent dehydration. Similar behavior has been reported for electrodeposited Ni-containing thin films, where tensile residual stresses exceed the mechanical stability limit of the deposited layer \cite{Garcia2025,Malik2004,Huff2022}.

Films deposited in NH$_4$Cl display intermediate morphological characteristics, combining relatively dense and continuous surfaces with moderate structural distortion. Although local coordination heterogeneity remains evident, particularly for NiHCF, these films preserve good morphological integrity, illustrating that local disorder does not necessarily compromise long-range structural continuity.

LiCl-derived FeHCF exhibits a compact morphology with closely packed domains and reduced porosity. This behavior is consistent with limited Li$^+$ incorporation arising from its strong hydration shell, which suppresses water co-insertion and framework growth \cite{Aniskevich2026,Li2018}. Nevertheless, additional cracking during SEM observation indicates that residual coordinated water remains sufficient to induce dehydration-related stress under vacuum.

As discussed previously, CoHCF and NiHCF could not be reproducibly electrodeposited in LiCl, further demonstrating that Li$^+$ does not support stable growth of these compositions under the investigated conditions.

\begin{figure}[h]
\centering
\includegraphics[width=\textwidth]{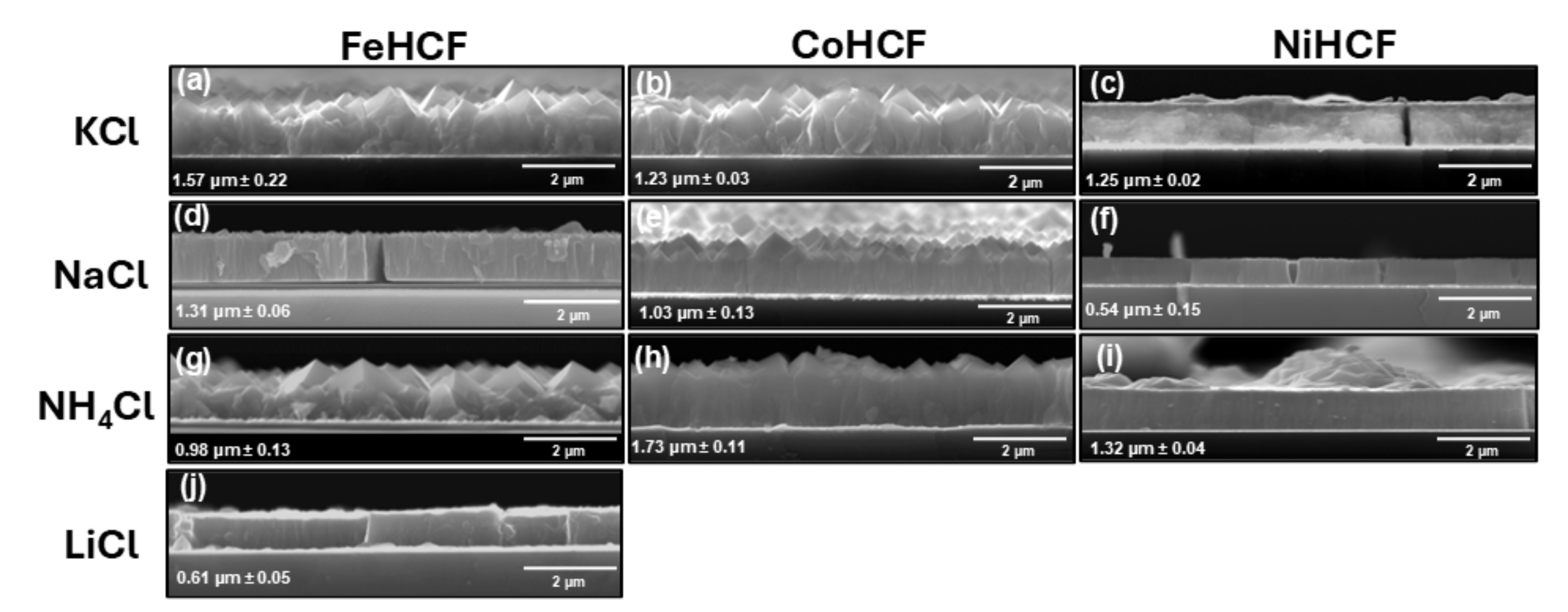}
\caption{
Cross-sectional SEM images of electrodeposited Prussian Blue analogue thin films illustrating the electrolyte-dependent film thickness and growth morphology. Variations in thickness, film continuity, and substrate coverage demonstrate the strong influence of the supporting electrolyte on film growth.
}
\label{fig:cross}
\end{figure}

Cross-sectional SEM analysis (Figure~\ref{fig:cross}) further reveals that film thickness and growth behavior are strongly dependent on the supporting electrolyte.

FeHCF forms relatively thick and continuous coatings in KCl, NaCl, and NH$_4$Cl, with thicknesses ranging from 0.98 to 1.57~$\mu$m (Table~\ref{tab:thickness}). CoHCF exhibits the strongest electrolyte dependence, reaching its maximum thickness in NH$_4$Cl ($1.73 \pm 0.11~\mu$m), whereas NiHCF generally forms thinner and less continuous films, particularly in NaCl ($0.54 \pm 0.15~\mu$m). The reduced thickness of NaCl-derived NiHCF, together with its extensive cracking, suggests that residual stress and structural disorder limit continuous film growth and compromise mechanical integrity \cite{Engwall2016}.

LiCl-derived FeHCF exhibits the thinnest FeHCF film ($0.61 \pm 0.05~\mu$m), consistent with the limited framework growth expected from the high desolvation barrier of Li$^+$.

\begin{table}[h]
\centering
\caption{Film thickness determined from cross-sectional SEM images.}
\begin{tabular}{lcc}
\hline
Electrolyte & Material & Thickness ($\mu$m) \\
\hline
KCl & FeHCF & $1.57 \pm 0.22$ \\
KCl & CoHCF & $1.23 \pm 0.03$ \\
KCl & NiHCF & $1.25 \pm 0.02$ \\
NaCl & FeHCF & $1.31 \pm 0.06$ \\
NaCl & CoHCF & $1.03 \pm 0.13$ \\
NaCl & NiHCF & $0.54 \pm 0.15$ \\
NH$_4$Cl & FeHCF & $0.98 \pm 0.13$ \\
NH$_4$Cl & CoHCF & $1.73 \pm 0.11$ \\
NH$_4$Cl & NiHCF & $1.32 \pm 0.04$ \\
LiCl & FeHCF & $0.61 \pm 0.05$ \\
\hline
\end{tabular}
\label{tab:thickness}
\end{table}

EDS analysis provides complementary compositional information that supports the structural characterization. Although EDS cannot determine oxidation states or quantify vacancy concentrations, it reveals clear electrolyte-dependent differences in elemental composition, oxygen content, and alkali incorporation.

Na-containing films exhibit measurable Na incorporation, indicating that interstitial alkali ions participate in framework charge compensation. Similar behavior was observed in our previous study on electrodeposited CoHCF and NiHCF, where increased K incorporation correlated with a higher Fe$^{2+}$/Fe$^{3+}$ ratio determined by XPS together with systematic changes in the CN stretching region \cite{Garcia2025}. These observations suggest that interstitial alkali ions modify the local electronic structure of the Fe--C$\equiv$N--M linkage.

The compositional trends closely parallel the structural characterization. NaCl-derived films exhibit the highest oxygen contents together with measurable Na incorporation, consistent with the greater lattice distortion, broader Raman bands, and extensive cracking observed by XRD, Raman spectroscopy, and SEM. In contrast, KCl-derived films show comparatively lower oxygen incorporation and a more homogeneous elemental distribution, in agreement with their higher crystallinity and improved structural coherence. NH$_4$Cl-derived films remain closer to the expected framework stoichiometry and generally exhibit intermediate oxygen contents, although the persistence of multiple Raman components, particularly for NiHCF, indicates that local coordination heterogeneity remains despite the comparatively ordered average crystal structure.

LiCl-derived FeHCF exhibits comparatively low oxygen incorporation together with the smallest film thickness and a compact morphology. These observations are consistent with limited Li$^+$ insertion resulting from its large hydration energy, leading to reduced framework growth and lower hydration than in NaCl-derived films.
\begin{table}[h]
\centering
\caption{
Atomic composition obtained by EDS mapping (top-view). 
\textit{Note:} NH$_4^+$ and Li$^+$ cannot be reliably quantified by EDS due to the overlap of nitrogen signals with CN ligands and the low sensitivity of the technique to light elements.
}
\label{tab:EDS_atomic}
\begin{tabular}{lccccccc}
\toprule
Material & C (\%) & N (\%) & O (\%) & Fe (\%) & Co (\%) & Ni (\%) & Electrolyte ion (\%) \\
\midrule

\multicolumn{8}{c}{\textbf{KCl}} \\
\midrule
FeHCF & 41.30 & 35.81 & 6.42 & 12.61 & -- & -- & K: 3.85 \\
CoHCF & 38.00 & 34.23 & 5.60 & 5.64 & 6.00 & -- & K: 10.31 \\
NiHCF & 43.73 & 38.10 & 1.84 & 4.42 & -- & 4.98 & K: 7.00 \\

\midrule
\multicolumn{8}{c}{\textbf{NaCl}} \\
\midrule
FeHCF & 38.05 & 38.41 & 15.54 & 6.79 & -- & -- & Na: 1.21 \\
CoHCF & 41.49 & 29.08 & 10.53 & 6.46 & 7.02 & -- & Na: 5.42 \\
NiHCF & 47.89 & 33.33 & 4.18 & 4.29 & -- & 6.35 & Na: 3.96 \\

\midrule
\multicolumn{8}{c}{\textbf{NH$_4$Cl}} \\
\midrule
FeHCF & 38.35 & 47.12 & 9.04 & 5.49 & -- & -- & -- \\
CoHCF & 37.14 & 44.57 & 6.55 & 5.67 & 6.07 & -- & -- \\
NiHCF & 41.48 & 51.38 & 3.87 & 1.57 & -- & 1.71 & -- \\

\midrule
\multicolumn{8}{c}{\textbf{LiCl}} \\
\midrule
FeHCF & 41.77 & 46.45 & 6.64 & 5.14 & -- & -- & -- \\

\bottomrule
\end{tabular}
\end{table}

Overall, SEM and EDS complement the diffraction and spectroscopic analyses by demonstrating that electrolyte composition influences not only crystallographic order but also film growth, hydration, and elemental composition. These structural modifications evolve across multiple length scales and collectively determine the morphology, structural coherence, and transport accessibility of electrodeposited PBA thin films.

\subsection{Electrochemical Behavior and Transport Kinetics}

Cyclic voltammetry was first performed during \textit{in situ} electrodeposition in electrolyte solutions containing 0.25 mM K$_3$[Fe(CN)$_6$, 0.25 mM metal precursor (Fe$^{3+}$, Co$^{2+}$, or Ni$^{2+}$), and 1 M supporting electrolyte (KCl, NaCl, LiCl, or NH$_4$Cl). The voltammetric response confirms that the observed faradaic processes originate from the formation and redox activity of the PBA films, while the Au/Cr/Si substrate contributes negligibly within the investigated potential window.

In Prussian Blue analogues, the electrochemical response is governed by the reversible Fe$^{2+}$/Fe$^{3+}$ redox process, which is coupled to the insertion and extraction of charge-compensating cations within the open cyanide framework. The accessibility of these redox-active sites depends not only on the intrinsic redox chemistry but also on the structural integrity of the Fe--C$\equiv$N--M network, since defects, coordinated water, and local structural disorder can influence ion diffusion and charge-transfer kinetics \cite{Fu20262,Fu20261}.

\begin{figure}[h]
\centering
\includegraphics[width=\textwidth]{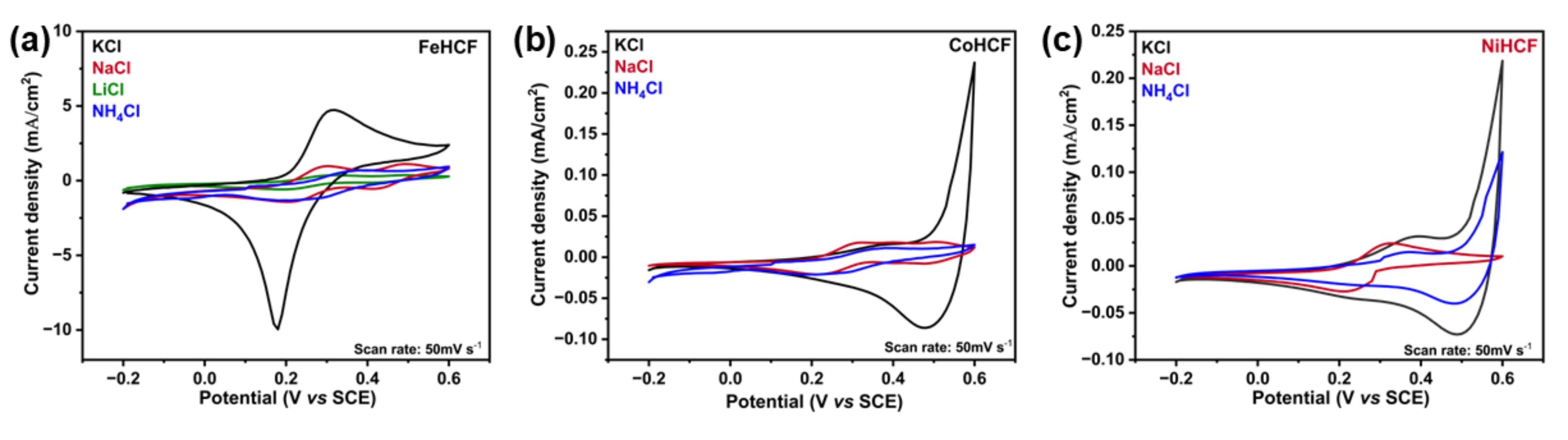}
\caption{
Cyclic voltammograms recorded during \textit{in situ} electrodeposition of FeHCF, CoHCF, and NiHCF in electrolytes containing 0.25 mM K$_3$[Fe(CN)$_6$, 0.25 mM metal precursor, and 1 M supporting electrolyte. Electrolyte-dependent variations in peak current and peak separation reflect differences in ion insertion and charge-transfer kinetics.
}
\label{fig:cv_electrolytes}
\end{figure}

Figure~\ref{fig:cv_electrolytes} shows that the electrochemical response depends strongly on the supporting electrolyte. FeHCF exhibits well-defined Fe$^{2+}$/Fe$^{3+}$ redox peaks whose current density, reversibility, and peak separation vary markedly with electrolyte composition.

Among the investigated electrolytes, KCl produces the sharpest voltammetric features, the highest peak currents, and the smallest peak separation, indicating relatively fast charge-transfer kinetics and efficient ion insertion into the PBA framework. Such behavior is consistent with a structurally coherent cyanide framework that facilitates reversible redox reactions and ion diffusion \cite{kjeldahl2019cation,liu2021prussian,Fu20262}. LiCl produces broader peaks with significantly larger peak separation, consistent with slower kinetics caused by the high hydration energy of Li$^+$, which increases the desolvation barrier prior to insertion \cite{nakamoto2024cathode,Bie2022}. NaCl exhibits intermediate behavior, whereas NH$_4$Cl yields comparatively symmetric redox peaks with reduced polarization, suggesting that hydrogen-bond interactions modify the insertion process \cite{Song2018Joule,Gao2024}.

These electrochemical trends closely follow the structural evolution revealed by XRD, Raman spectroscopy, SEM, and EDS. KCl-derived films exhibit the highest crystallinity, narrower Raman bands, compact morphologies, and lower oxygen contents, whereas NaCl-derived films display increased structural disorder, lattice distortion, oxygen incorporation, and extensive cracking. Together, these observations indicate that electrochemical performance depends primarily on the continuity of ion-transport pathways and the accessibility of redox-active sites rather than on lattice expansion alone, in agreement with recent studies highlighting the close relationship between framework structure, ion diffusion, and electrochemical performance in PBAs \cite{Simonov2020,Bostrom2022,Fu20261}.

Because KCl produced the most structurally coherent films and the most reproducible electrochemical response, it was selected for the detailed transport studies presented below. Scan-rate analysis and electrochemical impedance spectroscopy were therefore performed on pre-deposited FeHCF, CoHCF, and NiHCF thin films (~200 nm) prepared and measured under identical 1 M KCl conditions, allowing comparison of the intrinsic transport behavior of the three frameworks.

\subsubsection{Scan-rate Analysis}

\begin{figure}[h]
\centering
\includegraphics[width=\textwidth]{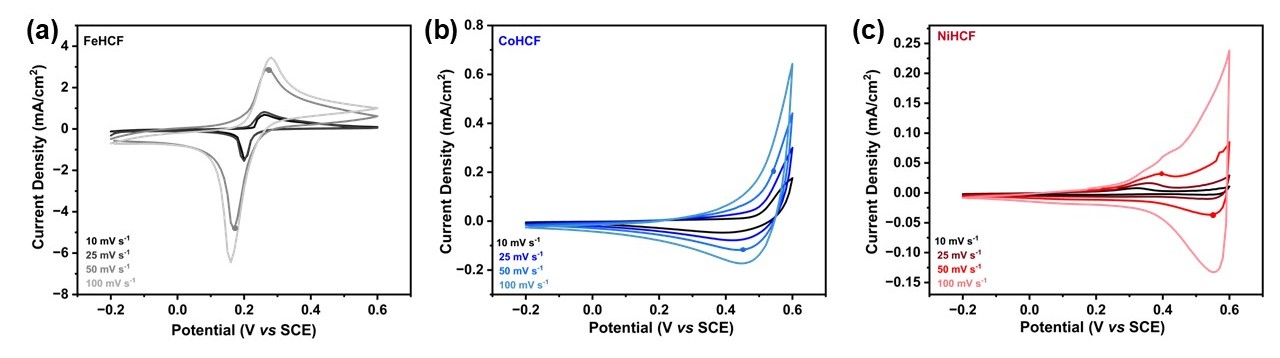}
\caption{Scan-rate-dependent cyclic voltammograms of pre-deposited PBA thin films ($\sim$200 nm) recorded in 1 M KCl at scan rates from 10 to 100 mV s$^{-1}$. The highlighted markers indicate the anodic and cathodic peak currents used to determine the apparent kinetic exponent ($b$) from the log($i$)--log($\nu$) analysis. The increase in current density with scan rate reflects the intrinsic transport kinetics of the different PBA frameworks.
}
\label{fig:scan_rate}
\end{figure}

\begin{figure}[h]
\centering
\includegraphics[width=0.75\textwidth]{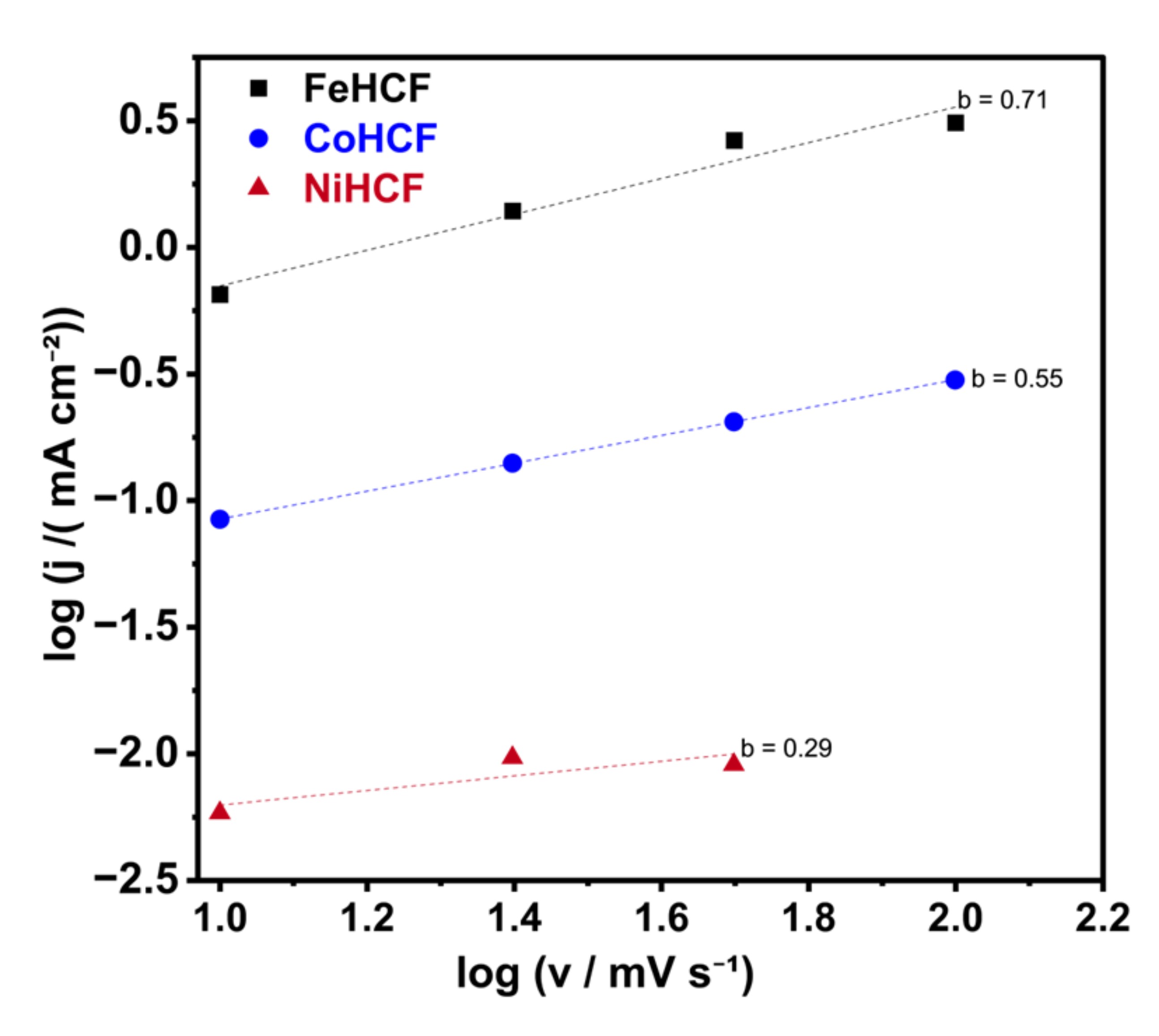}
\caption{
Log($i$)--log($\nu$) plots obtained from the scan-rate analysis of pre-deposited PBA thin films measured in 1 M KCl. The extracted apparent $b$-values indicate distinct transport regimes, ranging from mixed kinetic behavior for FeHCF ($b \approx 0.71$) to predominantly diffusion-influenced transport for CoHCF ($b \approx 0.55$) and increasingly non-ideal transport behavior for NiHCF ($b \approx 0.29$). The apparent $b$-values are interpreted here as empirical descriptors of transport behavior rather than rigorous kinetic exponents.
}
\label{fig:log_cv}
\end{figure}

To evaluate charge-storage kinetics, cyclic voltammograms were recorded between 10 and 100 mV s$^{-1}$ (Figure~\ref{fig:scan_rate}), and the corresponding log($i$)--log($\nu$) relationships were used to determine the apparent kinetic exponent ($b$) (Figure~\ref{fig:log_cv}).

The relationship between peak current and scan rate follows

\begin{equation}
i=a\nu^b,
\end{equation}

where $i$ is the peak current, $\nu$ the scan rate, and $b$ the apparent kinetic exponent. Under ideal conditions, $b\approx0.5$ corresponds to diffusion-controlled transport, whereas $b\approx1$ indicates predominantly surface-controlled charge storage. Since PBAs are structurally heterogeneous coordination frameworks, these assumptions are not strictly fulfilled. Consequently, the extracted $b$ values are interpreted as empirical descriptors of transport behavior rather than rigorous indicators of charge-storage mechanism \cite{Dunn2011,Augustyn2014}.

FeHCF exhibits the highest apparent exponent ($b\approx0.71$), indicating mixed surface-controlled and diffusion-influenced kinetics. CoHCF presents an intermediate response ($b\approx0.55$), whereas NiHCF exhibits a markedly lower apparent exponent ($b\approx0.29$), accompanied by broad, poorly defined redox peaks that progressively disappear at higher scan rates.

Rather than representing a conventional kinetic regime, the unusually low apparent $b$ value of NiHCF most likely reflects the breakdown of ideal power-law scaling under strongly heterogeneous transport conditions. The progressive decrease in $b$ from FeHCF to NiHCF therefore indicates increasing transport dispersion associated with structural disorder rather than a simple transition between diffusion- and surface-controlled processes.

This interpretation is fully consistent with the structural characterization. FeHCF exhibits the narrowest Raman bands and highest crystallographic coherence together with the largest apparent $b$ value, whereas NiHCF combines broad Raman features, reduced crystallinity, and the lowest apparent exponent. The same transport hierarchy is independently confirmed by impedance spectroscopy, where NiHCF exhibits both the highest transport resistance and the lowest CPE exponent.

Because all measurements were performed using identical thin films in the same electrolyte, the observed differences primarily reflect intrinsic framework-dependent transport properties. Electrolyte-specific kinetic effects therefore remain outside the scope of the present study.

Extended cycling was not performed because repeated scans progressively degraded the thin films, particularly at higher scan rates, leading to loss of voltammetric definition and partial film dissolution. Consequently, the present analysis focuses on comparing intrinsic transport behavior rather than evaluating long-term electrochemical stability or practical energy-storage performance.

\subsection{Impedance Response and Transport Limitations}

Electrochemical impedance spectroscopy (EIS) was employed to investigate the intrinsic transport properties of pre-deposited FeHCF, CoHCF, and NiHCF thin films prepared in 1 M KCl. To isolate framework-dependent transport, all measurements were performed under identical electrolyte conditions, minimizing contributions from differences in electrolyte conductivity, ion mobility, and interfacial effects.

Structural characterization was carried out on thicker films (approximately 1 $\mu$m) to ensure sufficient diffraction and Raman signal, whereas electrochemical measurements employed thinner films (approximately 200 nm) to improve electrochemical stability and reduce excessive transport resistance. Because all EIS measurements were performed using films of identical thickness, the discussion focuses on relative transport behavior rather than absolute impedance values.

\begin{figure}[htbp]
\centering
\includegraphics[width=\textwidth]{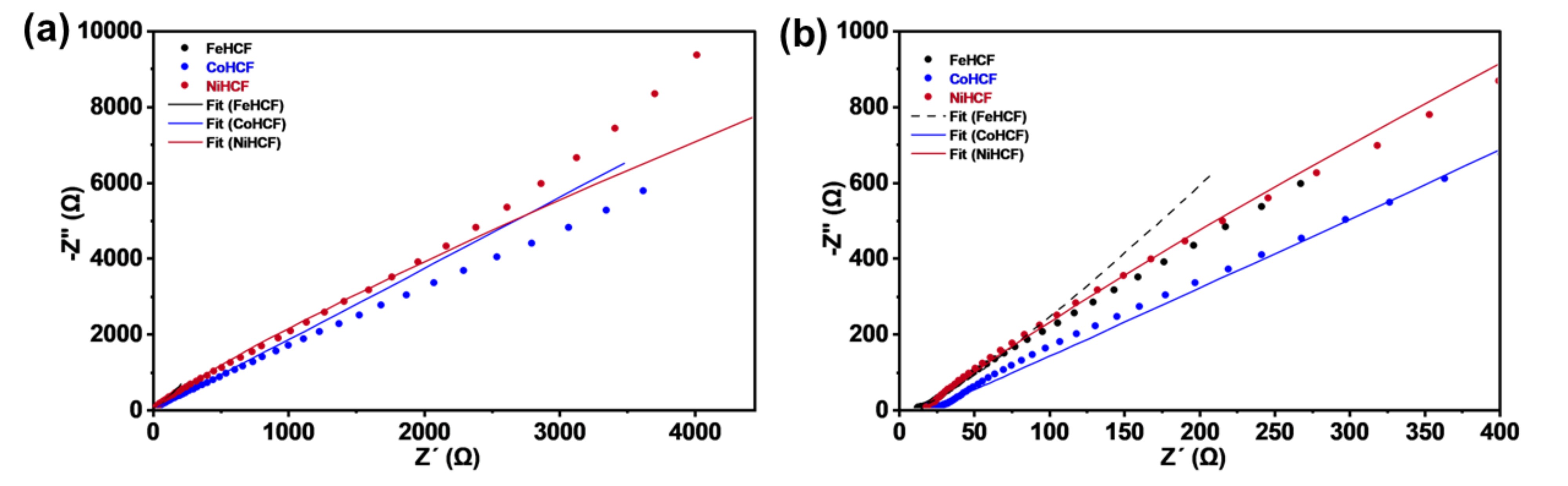}
\caption{
Nyquist plots of FeHCF, CoHCF, and NiHCF thin films measured in 1 M KCl: (a) full frequency range and (b) enlarged low-impedance region. The dashed line represents the ideal 45$^\circ$ response for semi-infinite diffusion, whereas the solid lines correspond to the equivalent-circuit fits.
}
\label{fig:eis}
\end{figure}

The Nyquist plots (Figure~\ref{fig:eis}) exhibit depressed impedance responses without a well-defined high-frequency semicircle, together with a quasi-linear low-frequency region. These characteristics indicate distributed ionic and electronic transport rather than a single charge-transfer process.

A systematic increase in impedance is observed across the FeHCF--CoHCF--NiHCF series. The low-frequency real impedance increases from approximately $2.6\times10^{2}$ $\Omega$ for FeHCF to nearly $9.0\times10^{3}$ $\Omega$ for NiHCF, corresponding to an increase of approximately one and a half orders of magnitude and indicating progressively restricted charge transport.

\begin{table}[htbp]
\centering
\caption{Characteristic impedance parameters obtained from the Nyquist plots.}
\label{tab:eis_summary}
\begin{tabular}{lccc}
\hline
Material & $R_s$ ($\Omega$) & $Z'$ at 0.1 Hz ($\Omega$) & Transport regime \\
\hline
FeHCF & 18.5 & $2.6\times10^{2}$ & Mixed \\
CoHCF & 18.7 & $3.6\times10^{3}$ & Diffusion-influenced \\
NiHCF & 17.8 & $9.0\times10^{3}$ & Strongly transport-limited \\
\hline
\end{tabular}
\end{table}

To quantify these trends, the spectra were fitted using the equivalent circuit
$R_s+(R||\mathrm{CPE})+W$ \cite{Brug1984}. Although more elaborate transmission-line models are commonly applied to porous electrodes, the compact geometry of the present thin films (approximately 200 nm) is adequately described by this simpler model while minimizing parameter correlation.

The fitted spectra reproduce the experimental response with relative fitting errors below 10\%, indicating satisfactory agreement between experiment and model. The extracted parameters should therefore be regarded as effective descriptors for comparing transport behavior rather than as a unique physical representation of the electrode/electrolyte interface.

\begin{table}[htbp]
\centering
\caption{Equivalent-circuit parameters obtained using $R_s+(R||\mathrm{CPE})+W$.}
\label{tab:eis_fit}
\begin{tabular}{lcccccc}
\hline
Sample & $R_s$ ($\Omega$) & $R$ ($\Omega$) & $Q$ (S s$^{n}$) & $n$ & $W$ ($\Omega$ s$^{-1/2}$) & $\chi^2$ \\
\hline
FeHCF & 18.0 & $1.3\times10^{2}$ & $1.5\times10^{-3}$ & 0.80 & $6.9\times10^{-4}$ & $1.5\times10^{-2}$ \\
CoHCF & 20.0 & $5.0\times10^{3}$ & $1.5\times10^{-4}$ & 0.72 & $5.0\times10^{-4}$ & $3.7\times10^{-2}$ \\
NiHCF & 18.1 & $8.9\times10^{3}$ & $5.0\times10^{-4}$ & 0.60 & $1.0\times10^{-3}$ & $1.4\times10^{-2}$ \\
\hline
\end{tabular}
\end{table}

The fitted solution resistance remains nearly constant ($R_s \approx 18$--20 $\Omega$), confirming that the electrolyte contribution is essentially identical for all measurements. In contrast, the transport resistance increases by almost two orders of magnitude from FeHCF to NiHCF, consistent with the progressive increase observed in the experimental spectra.

The constant phase element exponent decreases systematically from $n=0.80$ for FeHCF to $n=0.60$ for NiHCF, indicating increasingly non-ideal capacitive behavior. Because the CPE response reflects multiple contributions, including structural disorder, surface roughness, porosity, and non-uniform current distribution, the exponent is interpreted here as an empirical descriptor of transport heterogeneity rather than as evidence of a unique physical mechanism \cite{CordobaTorres2015,Hirschorn2010}.

Similarly, although the fitted Warburg element indicates diffusion-related transport limitations, the absence of an ideal 45$^\circ$ region shows that transport cannot be described by classical semi-infinite Fickian diffusion. Instead, the Warburg contribution is interpreted as an effective representation of distributed ion transport through a structurally heterogeneous framework \cite{Bisquert2000,Bisquert2002}.

The impedance results closely follow the structural trends established by XRD and Raman spectroscopy. FeHCF, which exhibits the highest crystallographic coherence and the narrowest Raman bands, also presents the lowest transport resistance and the largest CPE exponent. Conversely, NiHCF displays the broadest Raman features together with the highest impedance and the lowest $n$ value, consistent with increasingly dispersed transport pathways. This behavior agrees with the scan-rate analysis, where the apparent kinetic exponent decreases systematically from FeHCF to NiHCF.

Overall, the EIS results demonstrate that transport limitations are governed primarily by framework disorder and transport-pathway connectivity rather than by lattice expansion alone. Increasing structural heterogeneity broadens the distribution of transport timescales, reduces effective pathway connectivity, and progressively transforms the electrochemical response from relatively homogeneous transport in FeHCF to strongly distributed transport in NiHCF.

\subsection{Structure--Transport Correlation}

The combined XRD, Raman, cyclic voltammetry, and electrochemical impedance spectroscopy results establish a consistent relationship between structural disorder and electrochemical transport in electrodeposited Prussian Blue analogue (PBA) thin films. Although these techniques probe different length scales, they collectively show that increasing structural heterogeneity progressively degrades ion transport across the FeHCF--CoHCF--NiHCF series.

XRD reveals electrolyte-dependent variations in lattice parameter, peak broadening, and crystallinity, indicating changes in long-range structural order. Raman spectroscopy complements these observations by showing progressive broadening and splitting of the CN stretching modes, evidencing increasingly heterogeneous local coordination environments. Together, these results demonstrate that electrolyte composition governs both the average crystal structure and the local defect landscape of the PBA framework.

The electrochemical measurements closely follow these structural trends. FeHCF exhibits the lowest impedance together with the highest apparent kinetic ($b$) and CPE ($n$) exponents, whereas NiHCF displays the highest transport resistance and the lowest kinetic parameters. Rather than indicating distinct charge-storage mechanisms, these changes reflect progressively more heterogeneous ion transport associated with increasing structural disorder.

Importantly, the present results demonstrate that electrochemical performance cannot be explained solely by lattice expansion. Although a larger lattice parameter may facilitate ion insertion, it is accompanied here by increased structural disorder, including vacancies, coordinated water, local symmetry breaking, and microstrain, which reduce the continuity of transport pathways and increase transport resistance. Consequently, transport is governed primarily by the connectivity and structural coherence of the insertion network rather than by lattice dimensions alone.

Taken together, the results support a unified structure--transport relationship in which the supporting electrolyte controls defect formation during electrodeposition, thereby determining the connectivity of ion-transport pathways and the resulting electrochemical performance of PBA thin films.

\subsection{Mechanistic Interpretation of Electrolyte Effects}

The electrolyte-dependent structural evolution observed throughout this work can be interpreted in terms of the overall free-energy balance governing ion insertion into Prussian Blue analogues \cite{Marcus1994,Nightingale1959,Okoshi2013,Woodford2010}:

\begin{equation}
\Delta G_{\mathrm{total}}
=
\Delta G_{\mathrm{redox}}
+
\Delta G_{\mathrm{desolv}}
+
\Delta G_{\mathrm{electrostatic}}
+
\Delta G_{\mathrm{strain}}.
\end{equation}

Here, $\Delta G_{\mathrm{redox}}$ describes the intrinsic transition-metal redox process, $\Delta G_{\mathrm{desolv}}$ the energy required to remove the hydration shell of the inserting cation, $\Delta G_{\mathrm{electrostatic}}$ the interaction between the inserted ion and the negatively charged cyanide framework, and $\Delta G_{\mathrm{strain}}$ the energetic cost associated with lattice distortion, defect formation, hydration, and local structural rearrangement.

Although these contributions were not quantified experimentally, this framework provides a qualitative interpretation of the systematic trends observed across the investigated electrolytes.

The limited growth and electrochemical activity observed for LiCl-derived films are consistent with the large hydration energy of Li$^+$, which increases the desolvation barrier and hinders ion insertion into the PBA framework. Consequently, only FeHCF could be reproducibly deposited under the investigated conditions.

Na$^+$ partially overcomes this desolvation penalty but promotes a larger structural response upon insertion. The expanded lattice, broader Raman bands, higher oxygen content, and increased cracking observed for NaCl-derived films indicate that the energetic benefit of ion insertion is accompanied by a significant increase in structural strain and defect accommodation, resulting in a more heterogeneous framework.

Among the investigated electrolytes, K$^+$ provides the most favorable balance between desolvation, electrostatic stabilization, and structural accommodation. Consequently, KCl-derived films exhibit the highest crystallographic coherence together with the lowest transport resistance, indicating that structural distortion is minimized while maintaining efficient ion accessibility.

NH$_4^+$ exhibits distinct behavior because its interaction with the framework may involve hydrogen bonding in addition to electrostatic interactions. Although these interactions were not directly investigated, NH$_4$Cl-derived films combine relatively good crystallinity and compact morphology with evidence of local coordination heterogeneity, suggesting partial stabilization of the framework despite the persistence of local disorder.

Overall, these observations indicate that the supporting electrolyte acts as a defect-engineering parameter during electrodeposition. Rather than simply determining the identity of the charge-compensating ion, it controls the balance between desolvation, electrostatic stabilization, and lattice strain, thereby governing defect formation, structural heterogeneity, and ultimately the connectivity of ion-transport pathways within the PBA framework.

\section{Conclusions}

This work demonstrates that the supporting electrolyte is a key parameter governing the structural evolution and transport properties of electrodeposited Prussian Blue analogue (PBA) thin films. Systematic variation of the electrolyte cation (Li$^+$, Na$^+$, K$^+$, and NH$_4^+$) produced significant differences in crystallinity, local coordination, morphology, and electrochemical behavior across FeHCF, CoHCF, and NiHCF.

Combined XRD, Raman spectroscopy, SEM, EDS, cyclic voltammetry, and electrochemical impedance spectroscopy reveal that KCl promotes the formation of the most structurally coherent frameworks, whereas NaCl increases lattice distortion, hydration, defect formation, and mechanical degradation. LiCl limits framework growth because of the high desolvation energy of Li$^+$, while NH$_4$Cl yields comparatively coherent films despite retaining local coordination heterogeneity.

Electrochemical characterization reveals a direct relationship between structure and transport. Increasing structural disorder is accompanied by higher impedance, lower apparent kinetic exponents, and increasingly distributed transport behavior, indicating that transport limitations are governed primarily by defect-induced disruption of ion-transport pathways rather than by lattice expansion alone.

Overall, these findings demonstrate that electrolyte selection provides an effective strategy for defect engineering in electrodeposited PBA thin films and establish a mechanistic framework linking electrolyte chemistry, structural disorder, and electrochemical transport.

\clearpage

\bibliographystyle{unsrt}
\bibliography{cas-refs}

\section*{CRediT authorship contribution statement}
L. O. Garcia: Investigation, Writing – original draft.
M. Pohlitz: Methodology.
M. F. Kalady: Methodology.
C. K. Müller: Conceptualization, Supervision, Funding acquisition.

\section*{Declaration of competing interest}
The authors declare that they have no known competing financial interests.

\section*{Data availability}
The data that support the findings of this study are available from the corresponding author upon reasonable request.

\section*{Acknowledgments}
The authors thank Birgit Opitz from IFW Dresden for support with XRD measurements.

\section*{Funding}
This work was supported by the Deutsche Forschungsgemeinschaft (DFG, Project No. 531524052) and by the European Social Fund (ESF) and the Free State of Saxony within the Landesinnovationsförderprogramm (Application No. 100670500).


\end{document}